\documentclass[aps,prl,twocolumn,floatfix,preprintnumbers,nofootinbib]{revtex4-1}

\usepackage[margin=0.7in]{geometry}
\usepackage[utf8]{inputenc}

\usepackage{physics}
\usepackage{amsmath,amssymb,amsthm,amsfonts}
\usepackage[final]{graphicx}
\usepackage{subcaption}
\usepackage{lipsum}
\usepackage{mathtools}
\usepackage{enumitem,tabulary}
\usepackage{indentfirst}
\usepackage[english]{babel}
\usepackage{textcomp}
\usepackage{multirow}
\usepackage{tikz}
\usepackage{tipa}
\usepackage{CJKutf8}
\usepackage{feynmf}
\usepackage{slashed}
\usepackage{braket}
\usepackage[toc,page]{appendix}
\usepackage{url}
\usepackage{natbib}
\usepackage{graphicx}
\usepackage[export]{adjustbox}

\usepackage{mathrsfs}  
\usepackage{cancel}
\usepackage[normalem]{ulem}
\usepackage{array}
\usepackage{booktabs}
\usepackage{verbatim}
\usepackage{ragged2e}

\RequirePackage[colorlinks=true
,urlcolor=blue
,anchorcolor=blue
,citecolor=blue
,filecolor=blue
,linkcolor=blue
,menucolor=blue
,linktocpage=true
,pdfproducer=medialab
,pdfa=true
]{hyperref}

\begin{document}
\begin{flushright}
MI-HET-841
\end{flushright}

\title{Dark Matter Internal  Pair Production: A Novel Direct Detection Mechanism}

\author{Bhaskar Dutta}
\email{dutta@tamu.edu}
\affiliation{Department of Physics and Astronomy,
Texas A\&M University, College Station, Texas 77845 USA}

\author{Aparajitha Karthikeyan}
\email{aparajitha\_96@tamu.edu}
\affiliation{Department of Physics and Astronomy,
Texas A\&M University, College Station, Texas 77845 USA}

\author{Hyunyong Kim}
\email{hyunyong.kim@cern.ch}
\affiliation{Department of Physics and Astronomy, Seoul National University, Seoul 08826, Republic of Korea}

\affiliation{Department of Physics and Astronomy,
Texas A\&M University, College Station, Texas 77845 USA}

\author{Mudit Rai}
\email{muditrai@tamu.edu}
\affiliation{Department of Physics and Astronomy,
Texas A\&M University, College Station, TX-77845 USA}

\begin{abstract}
    We introduce a new dark matter detection mechanism, dark matter internal pair production (DIPP), to detect dark matter candidates at beam dump facilities. 
    When energetic dark matter scatters in a material, it can create a lepton-antilepton pair by exchanging a virtual photon with the nucleus, akin to the neutrino trident process. 
    We demonstrate this process for dark matter coupled to dark photons in experiments such as DarkQuest, SBND, and DUNE ND.
    Since the lepton-antilepton pair carries a significant fraction of dark matter energy, it can be clearly distinguished from backgrounds. 
    Utilizing the above features, we show that DIPP is highly effective at probing various dark matter models, particularly at DUNE ND and DarkQuest, by searching for electron-positron and muon-antimuon signatures. At DUNE ND, we find that the sensitivity of DIPP extends over a larger parameter space than the existing limits from electron and nuclear recoil.
    Additionally, we explore a scenario where dark sector couplings involve quarks and muons only, demonstrating that DIPP can probe a wide variety of dark matter models with different final states.
\end{abstract}

\maketitle

\textit{Introduction} - 
There is overwhelming evidence pointing toward the existence of dark matter (DM), which accounts for around 80\% of the matter content in the Universe~\cite{Freese:2017idy}, such as anomalous galactic rotation curves~\cite{babcock1939rotation}, x-ray mapping of bullet cluster mergers~\cite{Zwicky:1933gu, Zwicky:1937zza, Markevitch:2001ri, Markevitch:2004qk}, and structure formation simulations~\cite{Neto:2007vq, Maccio:2008pcd, Springel:2008cc}. While these establish that relic DM is cold and massive, its particle nature is yet to be determined. Weakly Interacting Massive Particles with masses of $\mathcal{O}(0.1-1~\text{TeV})$ have long been considered a leading candidate for explaining DM, where the relic abundance is naturally determined by annihilation freeze-out processes mediated by weak interactions. However, with the lack of a positive weakly interacting massive particle signatures~\cite{XENON:2018voc, XENON:2019gfn, XENON:2020fgj, DarkSide:2018ppu, SuperCDMS:2020aus}, 
there is a renewed interest in pursuing other possible explanations of DM.

One such explanation is sub-GeV DM~\cite{Boehm:2003bt, Boehm:2003hm} which interacts with the standard model (SM) sector via new dark sector portal particles or mediators~\cite{Fayet:2004bw, Huh:2007zw, Pospelov:2007mp, Chun:2010ve,Bi:2009uj,Park:2015gdo,Foldenauer:2018zrz,Dutta:2019fxn}. These DM particles can be searched at various energy and intensity frontier experiments where their probability of appearance and interaction depends on various SM particle energy fluxes, SM interaction strength of the mediators, etc. Many mechanisms have been employed to detect various sub-GeV DM interactions, primarily missing energy~\cite{Banerjee:2019pds, Andreev:2024lps, NA64:2024klw},  electron and nuclear recoils~\cite{LSND:2001akn,MiniBooNE:2017nqe,COHERENT:2021pvd, CCM:2021leg,Rott:2018rlw,SENSEI:2020dpa,Lattaud:2022jnq,PandaX:2024muv,XENON:2024wpa,SuperCDMS:2020aus,LZ:2023lvz,SuperCDMS:2024yiv}, nuclear deexcitation lines~\cite{Dutta:2023fij, Dutta:2024kuj}, and dark bremsstrahlung (dubbed as ``dark trident")~\cite{deGouvea:2018cfv, Adrian:2022nkt}. 

In this \textit{Letter}, we postulate a new mechanism called dark matter internal pair production (DIPP) and demonstrate its effectiveness in probing various sub-GeV DM models at proton fixed-target experiments. DIPP has a striking similarity with neutrino trident~\cite{Czyz:1964zz, Lovseth:1971vv, Fujikawa:1971nx, Brown:1971qr, Koike:1971tu, Brown:1972vne} as the internal legs of the Feynman diagrams of both processes include one portal mediator, one fermion, and one photon where the photon is exchanged between the lepton and a target nucleus. When DM passes through any material with enough kinetic energy, it can undergo DIPP, resulting in lepton-antilepton signatures. We find that for GeV scale energies with electron-positron and muon-antimuon final states, this process is coherently enhanced by the nuclear form factor~\cite{tsai:prim}, which is proportional to $Z^2$, $Z$ being the atomic number of the nucleus. 

DIPP is a powerful probe for sub-GeV DM for two main reasons. First, the lepton-antilepton final states of DIPP are less affected by neutrino or neutron-induced backgrounds at LArTPC~\cite{Gollapinni:2015lca}, or spectrometer~\cite{Apyan:2022tsd} detectors in comparison to single-electron or nuclear recoils.
Second, by analyzing the final states ($l^+ l^-$, for $l=e,\mu,\tau$ and/or hadronic $h^+ h^-$) DIPP can reveal the flavor or color-specific nature of dark sector models. DIPP thus broadens the scope of dark sector scenarios that can be examined. This is particularly advantageous compared to nuclear (electron) recoils, which only probe quark (electron) couplings. While dark tridents also point toward the model dependency, its cross section is suppressed by two off-shell massive mediators, whereas DIPP involves only one.

We will first introduce the vector-portal scalar DM models, followed by a brief discussion on their production in proton beam dump experiments. We then elaborate on DIPP's cross section to produce lepton-antilepton pairs. We analyze DIPP at experiments such as SBND~\cite{MicroBooNE:2015bmn}, DarkQuest Phase 1 and Phase 2~\cite{Apyan:2022tsd}, and DUNE Near Detector (DUNE ND)~\cite{DUNE:2021tad}. We explore kinematic cuts to mitigate backgrounds at these detectors. For DM coupled to dark photons, DIPP sensitivities at DarkQuest Phase 2 and DUNE ND probe uncharted parameter space. Additionally, for portals strongly coupled to quarks and muons, DIPP enables DUNE ND and DarkQuest Phase 2 to surpass current leading constraints from NA64$\mu$~\cite{NA64:2024klw}, and COHERENT CsI~\cite{COHERENT:2021pvd}.

\medskip

\textit{Dark matter models} - We first discuss scalar DM models with a vector portal to demonstrate DIPP. While fermionic DM scenarios are equally compelling and can also lead to DIPP (with a cross section twice that of scalar DM), scalar DM requires larger couplings to satisfy the relic abundance as compared to fermionic DM. Therefore, in light of the reach of upcoming experiments, we focus on scalar DM. The Lagrangian of scalar DM interacting with a general $U(1)_X$ mediator is
\begin{equation}
    \mathcal{L}_{\text{DM}} = \mathcal{L}^{\text{free}}_{Z'} + \mathcal{L}^{\text{free}}_{\chi} + i g_{Z'}Z'_{\mu}J^{\mu}_{f} + i g_{D}Z'_{\mu}J^{\mu}_{\chi}
\end{equation}
where $m_{\chi}$ and $m_{Z'}$ are the masses of $\chi$ and $Z'$ respectively. The scalar DM current is given by $J^{\mu}_{\chi} =
\chi^*(\partial^{\mu} \chi) -  (\partial^{\mu} \chi)^* \chi$. We assume that $Z'$ couples much more strongly to DM than to SM fermions, i.e. $g_{D} \gg g_{Z'}$ and that $Z'$s can kinematically decay into DM, i.e. $m_{Z'} > 2m_{\chi}$.
The SM current $J^{\mu}_f$ depends on the $U(1)_X$ model. To demonstrate DIPP, we consider two examples of $U(1)_X$ scenarios. The first is a $U(1)_X$ secluded model, where $Z' = A'$ is often referred to as a ``dark photon", couples to SM fermions with charges proportional to the $U(1)_{\text{em}}$ charges $q_f$, and $g_{Z'} = \epsilon e$. 
%Since it mixes with the SM photon ($g_{Z'} = \epsilon e$), the coupling to SM fermions is proportional to the $U(1)_{\text{em}}$ charges.
\begin{equation}
    J^{\mu}_{f} = J^{\mu}_{\text{em}} = \sum_f q_f \bar{f} \gamma^{\mu} f
\end{equation}

The second example consists of a gauge boson that predominantly couples to muons and quarks. For such scenarios, the SM current of interest can be expressed as

\begin{equation}
    J^{\mu}_{f} = \frac{g_B}{g_{Z'}}\sum_q \bar{q} \gamma^{\mu} q + \frac{g_\mu}{g_{Z'}}\bar{\mu} \gamma^{\mu} \mu
\end{equation} 

The above SM current appears in models such as $U(1)_{B-3L_\mu}$~\cite{Heeck:2018nzc, Han:2019zkz, Bauer:2020itv, delaVega:2021wpx},  $U(1)_{T_{3R}}$~\cite{Dutta:2019fxn, Dutta:2020jsy, Dutta:2020enk, Dutta:2022qvn} with right-handed couplings to muons and first-generation quarks, etc. This example aims to demonstrate DIPP's ability to probe DM portal interactions involving muons. However, selecting a model requires considering additional phenomenologies, such as those from neutrino couplings or loop-level interactions with electrons. Hence, we do not restrict ourselves to any particular muon or quarkphilic model in this Letter.

\textit{Dark Matter production} -
In a typical beam dump experiment, portal particles $Z'$s are first produced from neutral meson decays $\pi^0/ \eta \rightarrow \gamma Z'$~\cite{Batell:2009di} (predominantly for $m_{Z'} \lesssim 500$~MeV) and proton bremsstrahlung ~\cite{Foroughi-Abari:2021zbm}, $pN \rightarrow pN Z'$ ($m_{Z'} \gtrsim 500$~MeV) via their quark couplings. Once produced, they promptly decay into two DM particles with $\sim 100$\% probability.

Using Monte Carlo simulation techniques in Python, we simulate the energy-momentum fluxes of DM based on the probability distribution functions of various decay and scattering processes. Our simulation pipeline follows four main steps. First, we simulate the energy momentum of $Z'$s from the two-body decays of $\pi^0$ and $\eta$ mesons in the center-of-mass frame, using their differential branching ratios. We then boost them to the lab frame using the meson fluxes generated by \texttt{GEANT4}~\cite{GEANT4:2002zbu}. Since the proton energy is nearly constant, we calculate the differential cross section for proton bremsstrahlung in the laboratory frame to generate the kinematics and fluxes of $Z'$s. Secondly, we simulate the kinematics of the two DM candidates for each $Z'$ from the kinematics of a two-body decay. Third, from each DM momentum vector, we select the events that are within the solid angle subtended by the detector under consideration. Finally, we simulate lepton-antilepton signals produced via DIPP by integrating the DM flux with the exact DIPP differential cross section. In the following section, we will discuss the DIPP cross section for an incoming scalar SM to pair-produce $l^+ l^-$ via $Z'$ and our simulation technique.

\medskip

\begin{figure}[h]
    \centering
    \begin{subfigure}{.2\textwidth}
        \includegraphics[width=\textwidth, clip]{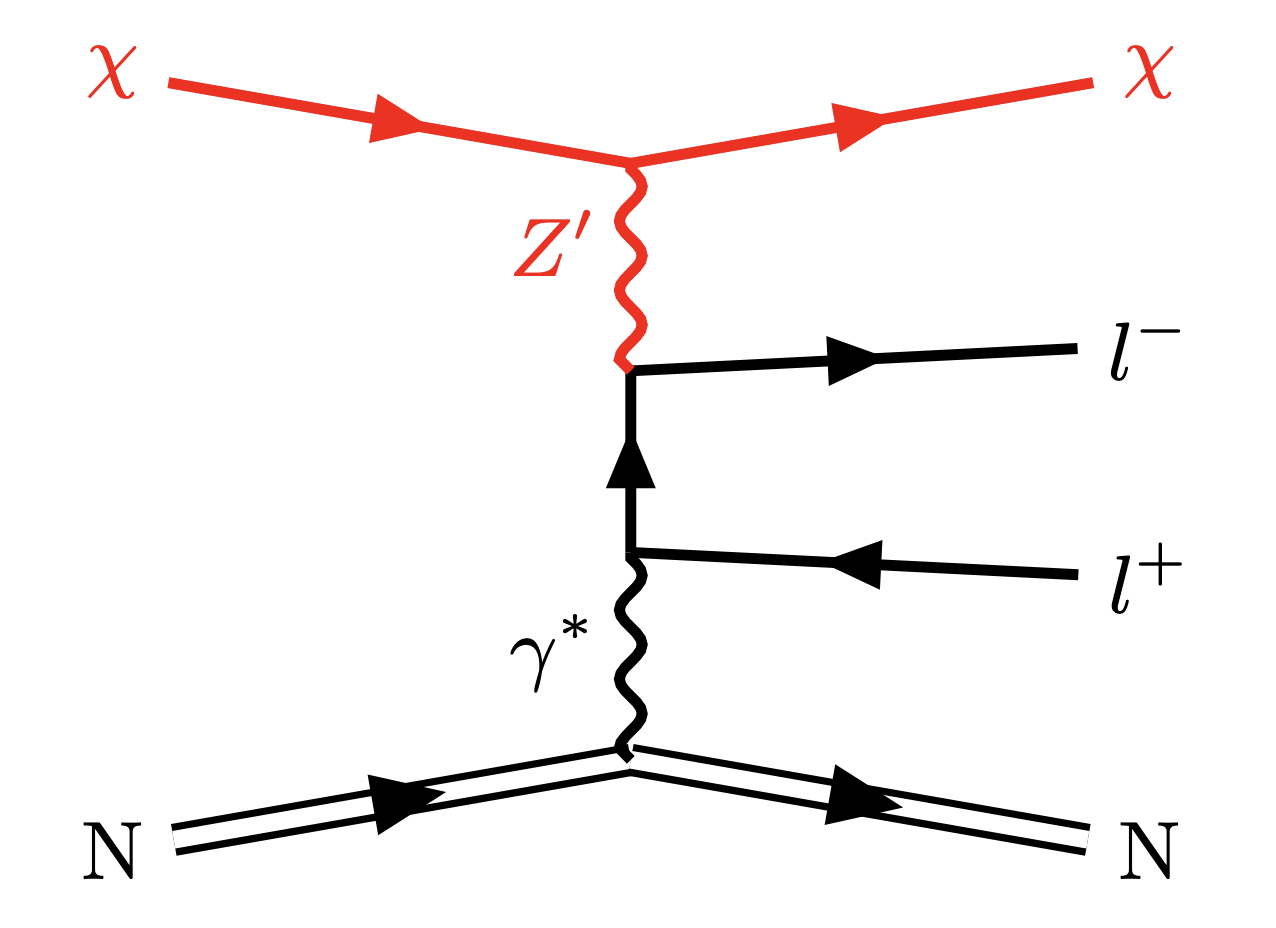}
        % \caption{}
    \end{subfigure}
    \hspace{0.1cm}
    \begin{subfigure}{.2\textwidth}
        \includegraphics[width=\textwidth, clip]{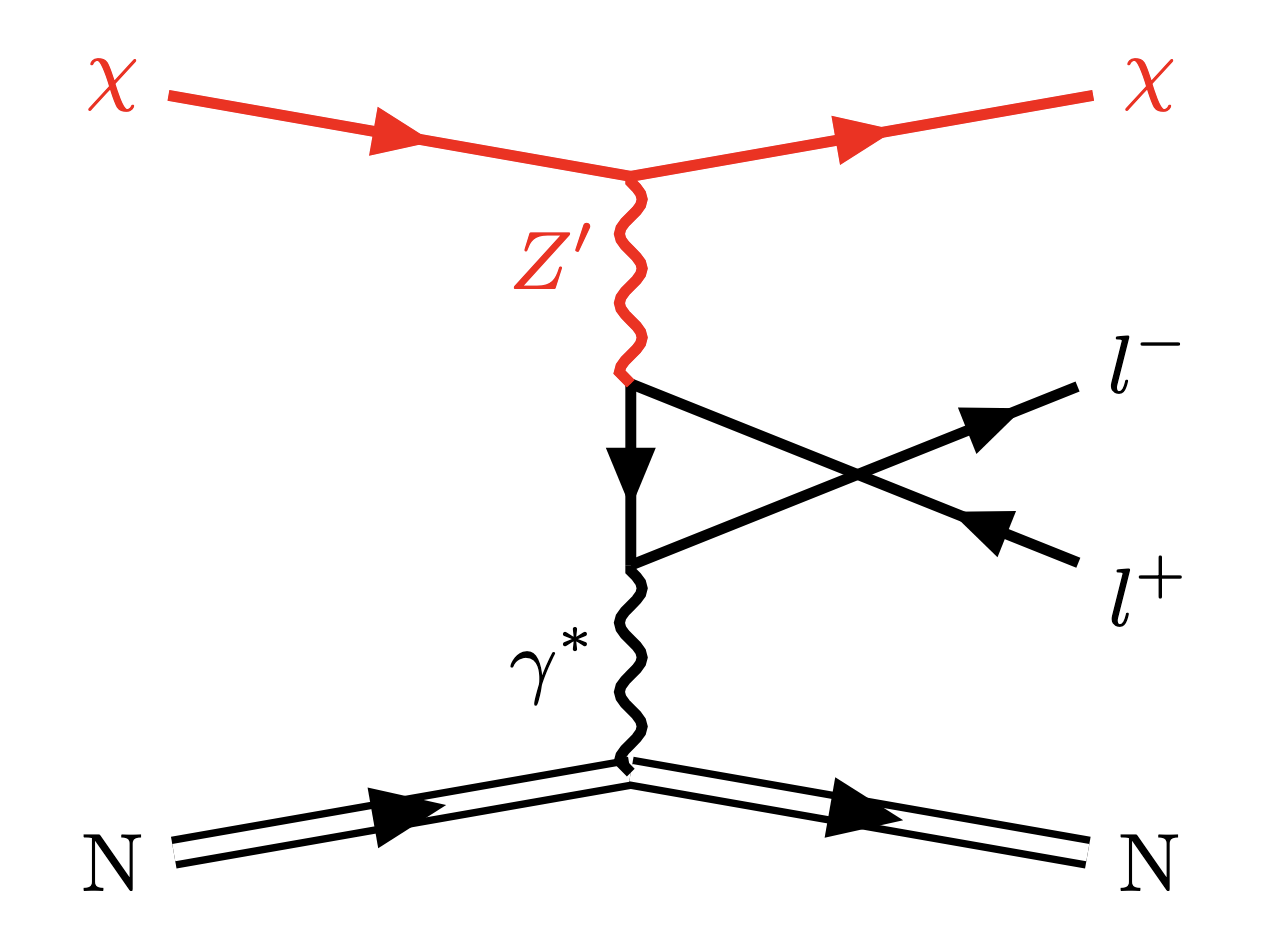}
        % \caption{}
    \end{subfigure}
    \captionsetup{justification=Justified}
    \caption{Feynman diagrams depicting dark matter internal pair production (DIPP) where DM candidate produces lepton-antilepton pair via new gauge boson $Z'$.}
    \label{fig:Tridentfeynman}
\end{figure}

\textit{DIPP cross section} -  
Figure~\ref{fig:Tridentfeynman} illustrates the DIPP process. Although this strongly resembles neutrino tridents, there are key differences such as (1) the Lorentz structure of DM, (2) the massiveness of DM, unlike nearly massless neutrinos, and (3) the presence of only one $Z'$, which mediates the interactions, while neutrino tridents in SM are mediated by $W$, $Z$ bosons, and potential new physics mediators.

We follow the prescription used for neutrino tridents in Refs.~\cite{Ballett:2018uuc, Altmannshofer:2019zhy, Zhou:2019vxt} to calculate the DIPP cross section. We first calculate the cross section of the 2-3 subprocess where DM scatters with the virtual photon exchanged between the lepton and the nucleus, i.e. $\chi/\chi^*(k_1) + \gamma^*(q) \rightarrow \chi/\chi^*(k_2) + l^-(p_2) + l^+(p_3)$.
This is expressed in terms of $Q^2 = -q^2$, and $\hat{s} = s - q^2$, where $s = (k_1 + q)^2$ is the center-of-mass energy of the 2-3 scattering process. We partition this cross section into transverse and longitudinal terms ($\sigma^{\text{T}}$, $\sigma^{\text{L}}$) corresponding to the photon polarities,

\begin{equation}
    \begin{aligned}
        \sigma^{\text{T}}_{\chi, \gamma}(Q^2, \hat{s}) = \frac{1}{2\hat{s}}\int \bigg(&-g^{\mu \nu} + \frac{4Q^2}{\hat{s}}k_1^{\mu}k_1^{\nu} \bigg) M_{\mu}M^*_{\nu}~d\text{PS}_3 \\
        \sigma^{\text{L}}_{\chi, \gamma}(Q^2, \hat{s}) = &\frac{1}{\hat{s}}\int \frac{4Q^2}{\hat{s}}k_1^{\mu}k_1^{\nu} M_{\mu}M^*_{\nu}~d\text{PS}_3
    \end{aligned}
    \label{eq:noEPA2to3cs}
\end{equation}

The phase space element $d\text{PS}_3$ is expressed in terms of five variables, two invariants $s_{23}, t_{a1}$, and three angles, $\theta'', \phi''$, and $\phi_R$. Further details on the matrix element that is calculated in \texttt{FeynCalc}~\cite{Shtabovenko:2016sxi,Shtabovenko:2020gxv, Mertig:1990an}, and the 2-3 phase space variables~\cite{Byckling:1971vca} can be found in Appendix~A. The differential cross section for the full 2-4 process is obtained by including the transverse and longitudinal components of the hadronic current~\cite{Ballett:2018uuc},

\begin{equation}
    \begin{aligned}
        \frac{d^2\sigma_{\chi A}}{dQ^2 d\hat{s}} = \frac{1}{32\pi^2}\frac{1}{\hat{s}Q^2} [ &h^{\text{T}}(Q^2, \hat{s})\sigma^{\text{T}}_{\chi, \gamma}(Q^2, \hat{s}) \\ &+ h^{\text{L}}(Q^2, \hat{s})\sigma^{\text{L}}_{\chi, \gamma}(Q^2, \hat{s})]
    \end{aligned}
    \label{eq:noEPAtotcs}
\end{equation}
%\end{widetext}
%where the limits for $\hat{s}$ and $Q^2$ depend on the regime of integration. 
In the coherent regime, the de-Broglie wavelength of the virtual photon is larger than the radius of the nucleus. Therefore, $Q_{\text{max}}(A) = \hbar c/(1.2\times A^{1/3}~\text{fm})$. Therefore, the limits for  $\hat{s}$ and $Q^2$ that satisfy coherence are,
\begin{equation}
    \begin{aligned}
        \hat{s} &: [(m_{\chi} + 2m_l)^2, m_{\chi}^2 + 2E_{\chi}Q_{\text{max}}(A)]\\
        Q^2 &: \bigg[ \bigg( \frac{\hat{s} - m_{\chi}^2}{2E_{\chi}} \bigg)^2, \min(Q_{\text{max}}(A)^2, \hat{s} - 2m_l^2) \bigg]
    \end{aligned}
    \label{eq:sq2limits}
\end{equation}

Using Monte Carlo methods, we generate samples of the seven variables that appear in the total differential cross section ($s_{23}, t_{a1}, \theta'', \phi'', \phi_R, \hat{s}, Q^2$).
%and include the appropriate Monte Carlo weights. 
As shown in Appendix~B, we determine the energy-momentum vector of the lepton-antilepton pair based on these seven variables. We then assign the differential cross section times the seven-dimensional Monte Carlo weight to each lepton-antilepton pair in the detector. The event kinematics and associated weights are used to determine the sensitivity of DIPP. We find that the events generated by our Monte Carlo method agree with those generated using \texttt{MadGraph5}~\cite{Alwall:2011uj}. The dependence of the total cross section on $E_{\chi}$, $m_{\chi}$, $m_{Z'}$, and $m_l$ is similar to those derived using the EPA approximation (as shown in Appendix~C). Since the validity of the EPA is not guaranteed for all our benchmark estimates, we use the total 2-4 cross section to predict event rates and sensitivities.

\medskip

\begin{figure}[h]
    %\centering
    \includegraphics[width=0.49\textwidth, left]{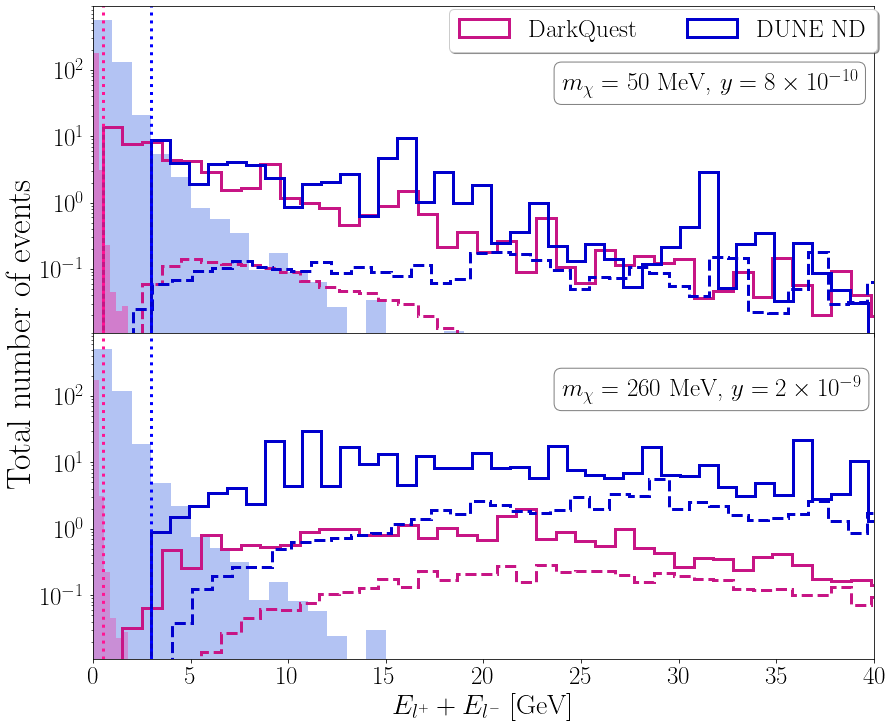}
    \captionsetup{justification=Justified, singlelinecheck=false}
    \caption{ Total energy of $e^+, e^-$ (solid) and $\mu^+ \mu^-$ (dashed) DIPP events that satisfy the kinematic cuts at DUNE ND (blue) and DarkQuest (dark pink). Shaded histograms depict the total backgrounds and the dotted vertical lines indicate the energy selection used at DUNE ND (blue) and DarkQuest (dark pink). The upper and lower panel corresponds to two different choices of mass and coupling parameters, ($m_{\chi}, y = \epsilon^2\alpha_D(m_{\chi}/m_{Z'})^4$): (50~MeV, $8\times 10^{-10}$) and (260~MeV, $2\times 10^{-9}$). }
    \label{fig:energyspectra}
\end{figure}

\begin{figure*}[t]
    \begin{subfigure}[t]{.46\textwidth}
        \includegraphics[width=\textwidth, clip]{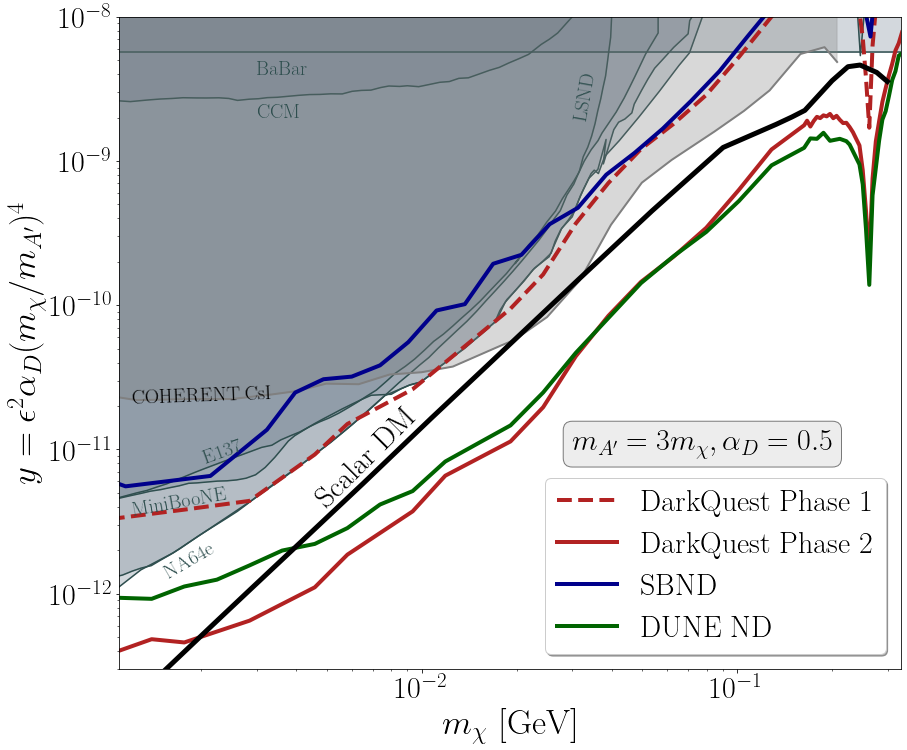}
        %\caption{}
        \label{fig:sensdp}
    \end{subfigure}
    \hspace{0.3cm}
    \begin{subfigure}[t]{.46\textwidth}
        \includegraphics[width=\textwidth, clip]{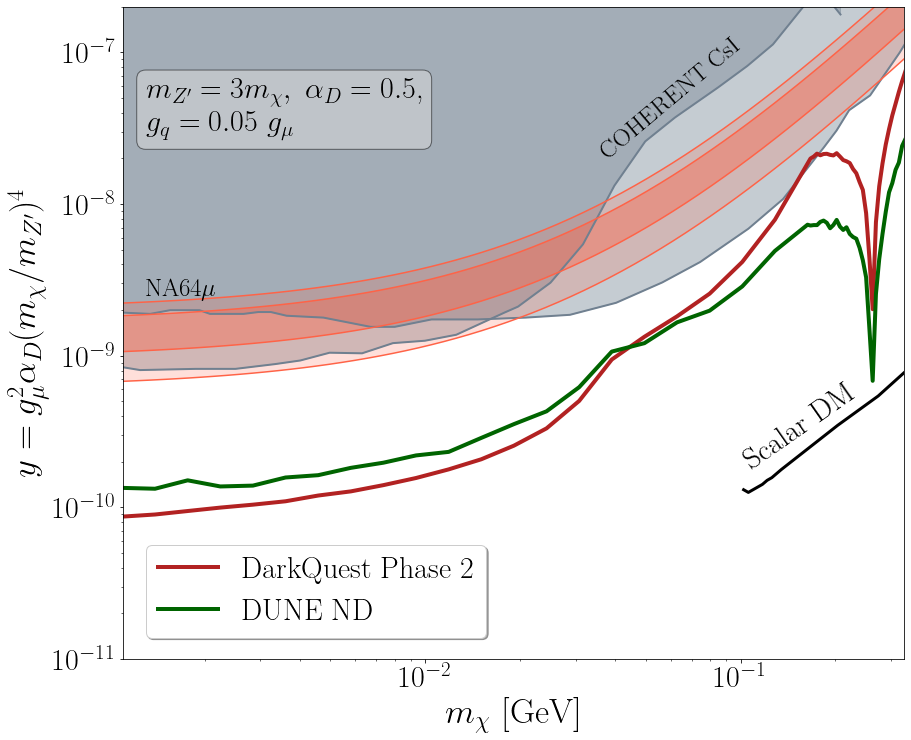}
        %\caption{}
        \label{fig:sensmuon}
    \end{subfigure}
    \captionsetup{justification=Justified, singlelinecheck=false}
    \caption{Sensitivity of DUNE ND, DarkQuest, and SBND to DIPP for (left) DM coupled to dark photons and (right) to $Z'$ with baryonic and muonic couplings only. }
    \label{fig:sensitivities}
\end{figure*}

\medskip

\textit{DIPP at benchmark experiments} - We examine DM sensitivities through DIPP in three benchmark experiments, SBND~\cite{MicroBooNE:2015bmn}, DUNE ND~\cite{DUNE:2021tad}, and DarkQuest~\cite{Apyan:2022tsd}. The ongoing SBND and future DUNE ND are LArTPC detectors, which are set to operate at the 8~GeV BNB beam~\cite{Machado:2019oxb}, and the 120~GeV LBNF beam~\cite{DUNE:2016hlj}, respectively. DarkQuest is a proton fixed-target beam dump spectrometer experiment that operates along the 120~GeV main injector beam~\cite{SeaQuest:2017kjt}. 

We consider a total of $1\times 10^{21}$ and $5.5\times 10^{21}$ POT for SBND and DUNE ND respectively. We estimate DarkQuest sensitivities for Phases 1 and 2 with $1.44 \times 10^{18}$ and $10^{20}$ POT, respectively. Our analysis includes a 2~m extension to the dump and the PbSc ECAL, placed $\sim 19$~m away from the target. Additionally, Phase 2 includes a hadronic absorber near the fourth tracking station.  

DM candidates at LArTPC DUNE ND have enough energy to produce $\mu^+ \mu^-$ and $e^+ e^-$ pairs via DIPP. However, these signals suffer from neutrino-induced backgrounds at the LArTPC detector. The main source of $\mu^+ \mu^-$ backgrounds is from $\nu_{\mu}$ CC, where pions are misidentified as muons. $e^+ e^-$ backgrounds mainly consist of $\nu_{\mu}-$NC $\pi^0$ events that decay into photons, where one of them is converted into $e^+e^-$. However, following the analysis in Ref.~\cite{Coloma:2023oxx} and using \texttt{GENIE}\cite{Andreopoulos:2009rq} generated events, we impose appropriate kinematic cuts to avoid neutrino-induced backgrounds. To avoid muon backgrounds, we select only those events at the detector where each of the muons and antimuons (1) has kinetic energy greater than 40~MeV, (2) $p_{T} < 125$~MeV, and (3) their angular separation is less than $40^0$.
For $e^+ e^-$ events, we consider only detector events that (1) have a reconstructed angle within $5^0$ from the beamline and (2) are separated between $1^\circ$ and $20^\circ$. Since the remaining events are soft, we achieve a background-free environment by having a (3) minimum reconstructed energy threshold of 3~GeV. More accurate predictions can be made after the experiment begins to run and we obtain the real backgrounds.

%for one event which includes a simultaneous hit of $\pi^{\pm} e^{\mp}$

Much like DUNE ND, DarkQuest is also sensitive to DM that undergoes both $e^+ e^-$ and $\mu^+ \mu^-$ DIPP at the ECAL and the hadronic absorber. While neutrino backgrounds from energetic $\pi^+$ and $K^+$ decays are strongly suppressed ($\sim10^{-8}$) by both the iron dump and the neutrino cross section, DarkQuest's main source of backgrounds is from $K^0_L$ decays, $K^0_L \rightarrow \pi^{\pm} + (e^{\mp} \nu_e)/(\mu^{\mp} \nu_{\mu})$ where the charged pion can be misidentified as an electron. These backgrounds are within $\mathcal{O}(1)$ for Phase 1, and $\mathcal{O}(100)$ for Phase 2.
Since these backgrounds have reconstructed energies lower than $\sim500$~MeV)~\cite{Apyan:2022tsd}, we estimate background-free sensitivities by requiring that each lepton has an energy greater than 250~MeV.

Figure.~\ref{fig:energyspectra} demonstrates the reconstructed energy spectra of the signal versus the background, where the stepped histograms show the $e^+e^-$ (solid) and $\mu^+\mu^-$ (dashed) spectra, and the shaded histograms show the background spectra, both DUNE ND and DarkQuest. We observe that the lepton-antilepton signals from DIPP at these $120$~GeV facilities are notably more energetic than the backgrounds.

SBND, operating at the 8~GeV facility, is more sensitive to $e^+ e^-$ DIPP events than $\mu^+ \mu^-$. Although the $e^+ e^-$ backgrounds, from $\nu_\mu$-NC~$\pi^0$ events, are subdominant compared to $\mu^+ \mu^-$, they can be further reduced by utilizing the difference in the arrival time of DM~\cite{SBND:2024vgn, bsmsbnd, bsmsbnd2} versus neutrinos. We anticipate a starker difference by studying those DM signals produced from $Z'$s at the dump (not in this study). Therefore, of the $19$~ ns between two proton spills, we expect negligible backgrounds in the later $\sim9$~ns\footnote{Private communication with V. Pandey from SBND}. More details on the timing spectra are in Appendix~D.

\medskip

\textit{Results} -  
Figure.~\ref{fig:sensitivities} (left) shows the DIPP sensitivity of DM coupled to dark photons at DUNE ND, DarkQuest, and SBND for $m_{A'} = 3 m_{\chi}$, and $\alpha_D = g_D^2/4\pi = 0.5$. The cross sections along the black line correspond to those in which scalar DM freezes out in agreement with the observed relic abundance~\cite{komatsu2009five}. The gray-shaded regions are excluded from electron recoil searches in LSND~\cite{LSND:2001akn}, MiniBooNE~\cite{MiniBooNE:2017nqe}, NA64e~\cite{Banerjee:2019pds}, and E137~\cite{Batell:2014mga},  $l^+ l^-$ searches in BABAR~\cite{BaBar:2008aby}, nuclear recoil searches in CCM~\cite{CCM:2021leg} and COHERENT~\cite{COHERENT:2021pvd}.
We find that DUNE ND and DarkQuest Phase 2 can explore new parameter spaces through DIPP that are consistent with the scalar DM relic density. Since DIPP-induced events are much more energetic than electron recoil events~\cite{Breitbach:2021gvv}, they suffer less from the backgrounds and hence dominate over electron-recoil. In contrast, SBND, with its lower energy, and DarkQuest Phase 1, with its lower luminosity, do not probe any new regions via DIPP. Since even the background-free SBND sensitivity does not probe new parameter space, we refrain from a detailed background analysis in this study. The sharp peak at $m_{\chi} \sim 250$~MeV ($m_{Z'} \sim 750$~MeV) is due to the enhancement of the $\rho$ meson in $A'$ production by proton bremsstrahlung.

Because of the proximity of the ECAL to the target ($\sim 20$~m), DarkQuest has a higher efficiency per POT than DUNE ND for DM produced from $\pi^0$ and $\eta$ decays. Therefore, the sensitivity of DarkQuest (Phase 2) is more enhanced for lighter masses $m_{Z'} \lesssim 500$ MeV ($m_{\chi} \lesssim 160$ MeV). 
Since heavier DM is predominantly produced from the forward-boosted proton bremsstrahlung, both detectors have similar efficiencies. Thus, factoring in total POT and backgrounds, DarkQuest Phase 2 is more sensitive for $m_{\chi} \lesssim 70$ MeV, while DUNE ND dominates at higher masses. Because of the DIPP's dependence on energy and mass scales, $e^+ e^-$ signals dominate over $\mu^+ \mu^-$ for lighter DM ($m_{\chi} \lesssim 100$~MeV) and are comparable for heavier DM.

A comparison of the DIPP sensitivities with DM electron recoil and dark trident is shown in Fig.~\ref{fig:DIPP_DT_DE} of Appendix~E. Here, we observe the superiority of DIPP over dark trident due to its larger cross section and over electron recoils due to its harder energy spectra. The sensitivities would be slightly altered with larger energy thresholds (for example,  5 (1)~GeV threshold at DUNE ND (DarkQuest) instead of 3(0.5)~GeV), which shows up as $\lesssim 10\%$ change in the sensitivity for lower masses.Figure.~\ref{fig:DIPP_DT_DE} also shows constraints from direct detection experiments like SENSEI~\cite{SENSEI:2024yyt}, XENON1T~\cite{XENON:2019zpr}, and PandaX~\cite{PandaX:2023xgl}, as well as projections from Belle~II~\cite{Inguglia:2019gub} and LDMX phase I~\cite{LDMX:2018cma}. DIPP sensitivity, in conjunction with other searches, complements them to provide broader insights into various models.

DUNE ND and DarkQuest can test a combination of the mediator's coupling to both quarks and muons, with $N_{\text{DIPP}} \propto g_q^2 g_\mu^2 g_{D}^2$ via DIPP. Therefore, DIPP offers a unique advantage compared to existing searches such as nuclear recoils (NR) at COHERENT and CCM that probe only the quark couplings ($N_{\text{NR}} \propto g_q^4 g_\chi^2$), and missing energy searches at NA64$\mu$, probing only the muon couplings with $N_{\text{missing}} \propto g_{\mu}^2$. Hence, DIPP can set limits that are complementary to the bounds from COHERENT and NA64$\mu$. We illustrate this feature with another example where the portal gauge boson exclusive couples to muons and first-generation quarks. Here, DUNE ND and DarkQuest Phase 2 are sensitive to DM scattering via muon-antimuon DIPP. Figure~\ref{fig:sensitivities} (right) depicts the sensitivity for $g_q/g_\mu = 0.05$, $m_{Z'} = 3m_{\chi}$, and $\alpha_D = 0.5$. The black contour represents parameters satisfying the relic abundance, where DM dominantly annihilates into $\mu^+ \mu^-$. It is also important to note that, based on the model specifics and the $Z'-\gamma$ mixing, additional constraints and relic abundance contributions from loop-suppressed electron couplings can be incorporated. The sensitivity of DUNE ND and DarkQuest Phase 2 to muon-antimuon states delves into new parameter space beyond the current limits from COHERENT~\cite{COHERENT:2021pvd}, NA64$\mu$~\cite{NA64:2024klw}, $(g-2)_\mu$~\cite{Muong-2:2021ojo, Muong-2:2021vma}. We see that DIPP can probe unexplored parameters for a large range of models with coupling ratios ranging between: $0.01 < g_q/g_\mu < 0.5$.

\medskip

\textit{Conclusions} - We introduce dark matter internal pair production (DIPP) as a novel detection method for DM that interacts with SM particles via portals. We outline DIPP's cross section and discuss our simulation methodology used for the full 2-4 scattering process. 
We identify the lepton-antilepton signatures from DIPP while accounting for SM backgrounds at three benchmark experiments using kinematic cuts. We illustrate DIPP for DM coupled to dark photons at DUNE ND, DarkQuest, and SBND, examining electron-positron and muon-antimuon final states. DIPP allows us to probe new parameter spaces consistent with thermal relic abundance for scalar DM. Because of the energetic nature of the signals, the DIPP analysis at DUNE ND probes a much larger parameter space than electron recoil. We also find that DIPP's cross section is larger than that of dark trident for the range of couplings that probe the relic abundance, and hence more favorable. We also study vector mediator interactions with quarks and muons at DUNE ND and DarkQuest, finding muon-antimuon detection complementary to NA64$\mu$'s missing energy and COHERENT's nuclear recoil searches. 
%Additionally, by combining these experiments and final states, we can probe electron and quark couplings for light mediators.

DIPP's ability to probe a wide range of DM models can be utilized to look for other scenarios such as $U(1)_{B-L}$, $U(1)_{L_i-L_j}$, etc. One can also look for unique tau lepton signatures~\cite{Altmannshofer:2024hqd, Bigaran:2024zxk} by probing DIPP in energy frontier experiments. We can extend this analysis beyond lab-based experiments to probe ambient DM at large volume $\nu$ detectors like DUNE FD, HyperK, JUNO, Borexino, etc.

\textit{Acknowledgements} - 
We thank P.S. Bhupal Dev, Brian Batell, Wooyoung Jang, Kevin J. Kelly, Doojin Kim, Vishvas Pandey, Zahra Tabrizi, and Adrian Thompson for their useful discussions. We especially thank A. Thompson for the background-related discussions. The work of B.D., A.K., H.K., and M.R. is supported by the U.S.~Department of Energy grant no. DE-SC0010813. 

\appendix

%%%%%%%%%%%%%%%%%%%%%%%%%%%%%%%%%%%
%\onecolumngrid
\medskip
\section{END MATTER}
\textit{Appendix A: 2-3 scattering} - The matrix element for the 2-3 sub-process $\chi \gamma \rightarrow \chi l^- l^+$ is

\begin{widetext}
    \begin{equation}
        \mathcal{M}^{\alpha} = (k_{1}^{\mu} + k_{2}^{\mu}) \bigg( \frac{1}{t_{a1} - m_{Z'}^2}\bigg) \bar{u}(p_2) \bigg[\gamma_{\mu} \big( g^l_V + g^l_A \gamma^5\big) \frac{(\cancel{q} - \cancel{p}_3 + m_{l})}{(q - p_3)^2 - m_{l}^2}\gamma^{\alpha} + 
        \gamma^{\alpha}\frac{(\cancel{p}_2 - \cancel{q} + m_{l})}{(p_2 - q)^2 - m_{l}^2}\gamma_{\mu} \big( g^l_V + g^l_A \gamma^5\big) \bigg] v(p_3) 
    \end{equation}
    \label{eq:matel}
\end{widetext}

The minimum energy required for a DM with mass $m_{\chi}$ to produce a lepton-antilepton pair (with lepton mass $m_l$) while interacting with a material whose nucleus has a mass $m_N$ can be expressed as follows,
\begin{equation}
    E_{\chi}^{\text{min}} = m_\chi + 2m_l \bigg( 1 + \frac{m_\chi + m_l}{m_N} \bigg)
\end{equation}

The square of the matrix element can be written in terms of the following mass invariants $s = (p_a + p_b)^2, s_{23} = (p_2 + p_3)^2, t_{a1} = (p_a - p_1)^2, t_{b3} = (p_b - p_3)^2,$ and $s_{12} = (p_1 + p_2)^2$. One can express $t_{b3}$ and $s_{12}$ in terms of the polar and azimuthal angles of $p_3$, $\theta''$ and $\phi''$, in the rest frame of particles 2 and 3, where $\Vec{p_{23}} = 0$. The $z$ axis is aligned along $\Vec{p_b}$ in the rest frame of particles 2 and 3. The 2-3 phase factor for a fixed $s$ is,
\begin{equation}
    d\text{PS}_3 = \frac{1}{2}\frac{1}{(4\pi)^2}\sqrt{1 - \frac{4m_l^2}{s_{23}}}\frac{dt_{a1}}{2\sqrt{\hat{s}^2 + 4m_{\chi}^2Q^2}}\frac{ds_{23}}{2\pi}\frac{d\Omega''}{4\pi}\frac{d\phi_R}{2\pi}
\end{equation}
The angle $\phi_R$ specifies the plane on which the 2-3 process occurs and is thus integrated between 0 and $2\pi$. The limits of the Mandelstam $s_{23}$ are $[4m_l^2, ~(\sqrt{\hat{s} - Q^2} - m_{\chi})^2]$ and those for $t_{a1}$ are
\begin{equation}
    \bigg[ 2m_{N}^2 - \frac{(\alpha_1 + \alpha_2)}{2(\hat{s} - Q^2)},~ 2m_{N}^2 - \frac{(\alpha_1 - \alpha_2)}{2(\hat{s} - Q^2)} \bigg]
\end{equation}

where
\begin{equation}
    \begin{aligned}
        \alpha_1 &= (m_{\chi}^2 + \hat{s})(m_{\chi}^2 - Q^2 - s_{23} + \hat{s}) \\
        \alpha_2 &= \sqrt{(4m_{\chi}^2Q^2 + (m_{\chi}^2 - \hat{s})^2)}\\ &\times \sqrt{((m_{\chi}^2 + Q^2 + s_{23} - \hat{s})^2 
        + 4m_{\chi}^2 s_{23})} 
    \end{aligned}
\end{equation}
    
The solid angle $d\Omega'' = d\cos{\theta''}d\phi''$ is integrated between $-1 \leq \cos{\theta''} \leq 1$ and $0 \leq \phi'' \leq 2\pi$.  

\textit{Appendix B: Lepton Kinematics from DIPP} - Following the prescription of Ref.~\cite{Zhou:2019vxt}, we replace the neutrino with a massive scalar DM candidate. In the CM frame of the $\chi$–A interaction:

\begin{equation}
    \begin{aligned}
        k^c_1 &= \left( \frac{s + Q^2 + m_{\chi}^2}{2\sqrt{s}}, ~p_i (\sin\theta, 0, -\cos\theta) \right)\\
        q^c &= \left( \frac{s - Q^2 - m_{\chi}^2}{2\sqrt{s}}, ~p_i (-\sin\theta, 0, \cos\theta) \right)\\
        k^c_2 &= \left( \frac{s - s_{23} + m_{\chi}^2}{2\sqrt{s}}, 0, 0, -p_f \right) \\
        p^c &= \left( \frac{s + s_{23} - m_{\chi}^2}{2\sqrt{s}}, 0, 0, p_f \right)
    \end{aligned}
    \label{eq:kinCOM}
\end{equation}

where
\begin{equation}
    \begin{aligned}
        p_i &= \frac{\sqrt{(s - m_{\chi}^2 + Q^2)^2 + 4m_{\chi}^2Q^2}}{2\sqrt{s}},\\
        p_f &= \frac{\sqrt{(s - m_{\chi}^2 - s_{23})^2 - 4m_{\chi}^2 s_{23}}}{2\sqrt{s}}.
    \end{aligned}
    \label{eq:momenta}
\end{equation}

Defining $t = 2q \cdot (k_1 - k_2)$, we find
\begin{equation}
    t = 2 \left( \frac{(s - Q^2 - m_{\chi}^2)(s_{23} + Q^2)}{4s} + p_i^2 - p_i p_f \cos \theta \right),
    \label{eq:tcostheta}
\end{equation}
which yields
\begin{equation}
    \begin{aligned}
        \cos \theta &= \big[(s - Q^2 - m_{\chi}^2)(s_{23} + Q^2) \\
        &+ (s - m_{\chi}^2 + Q^2)^2 + 4m_{\chi}^2Q^2 - 2st \big]/(4sp_i p_f).
    \end{aligned}
    \label{eq:costhetat}
\end{equation}

This expression for $\cos \theta$ can be used in Eq.~\eqref{eq:kinCOM}. Since the angular variables in the phase space integral are defined in the rest frame of $s_{23}$, we boost to that frame using Lorentz factors (defined as T2):

\begin{equation}
    \begin{aligned}
        \gamma &= \frac{s + s_{23} - m_{\chi}^2}{2\sqrt{s s_{23}}}, \\
        \beta_z &= \frac{\sqrt{(s - m_{\chi}^2 - s_{23})^2 - 4m_{\chi}^2 s_{23}}}{s + s_{23} - m_{\chi}^2},\\
        \beta_x &= \beta_y = 0
    \end{aligned}
    \label{eq:T1}
\end{equation}

We then rotate to align $\vec{q}$ with the z axis, defining this rotation as R2,
\begin{equation}
    \sin \theta_p = \frac{q'^c_{x}}{\sqrt{q'^2_x + q'^2_z}}, \quad
    \cos \theta_p = \frac{q'^c_{z}}{\sqrt{q'^2_x + q'^2_z}}.
    \label{eq:R1}
\end{equation}

In this frame, the leptons have momenta:
\begin{equation}
    \begin{aligned}
        p''^c_2 &= \left( \frac{\sqrt{s_{23}}}{2}, \frac{-\sqrt{s_{23} - 4m_l^2}}{2} \hat{\Omega}'' \right), \\
        p''^c_3 &= \left( \frac{\sqrt{s_{23}}}{2}, \frac{\sqrt{s_{23} - 4m_l^2}}{2} \hat{\Omega}'' \right),
    \end{aligned}
\end{equation}
where $\hat{\Omega}'' = (\sin \theta'' \cos \phi'', \sin \theta'' \sin \phi'', \cos \theta'')$.

Through Monte Carlo sampling, we obtain $p''^c_{2,3}$ in this frame with $\vec{q} \parallel z$ and $\vec{p}_{23} = 0$. The momenta in the $\chi$–A center-of-momentum frame are given by,
\begin{equation}
    [p^c_2, p^c_3] = \text{T}2^{\dagger} \cdot \text{R}2^{\dagger}[p''^c_2, p''^c_3].
\end{equation}

To transform to the lab frame, we first rotate to align $k_1^c$ with the $z$-axis. Since $k_1^c$ subtends an angle $\pi - \theta$ with $p^c$, we apply a $y$-axis rotation,
\begin{equation}
    [\bar{k}_1, \bar{q}, \bar{k}_2, \bar{p}] = R_{y}^{\pi - \theta} [k_1^c, q^c, k_2^c, p^c].
\end{equation}

The boost back to the lab frame uses the initial four-momentum $p_1 + q$. Letting $p_N = (m_N, 0)$, $p'_N = (m_N + E_R, \vec{p}_R)$, and $q = (-E_R, -\vec{p}_R)$, we define

\begin{equation}
    \begin{aligned}
        \vec{p}_R &= p_R(\sin\theta_R \cos\phi_R, \sin\theta_R \sin\phi_R, \cos\theta_R),\\
        E_R &= Q^2 / (2m_N), \quad p_R = \sqrt{E_R^2 + 2m_N E_R},\\
        \cos \theta_R &= \big( \hat{s}/2E_{\chi} + E_R \big) /p_R.
    \end{aligned}
\end{equation}

Choosing $\phi_R \in [0, 2\pi]$ defines the boost vector
\begin{equation}
    \begin{aligned}
        -p_{\text{boost}} = p_1 + q = (E_{\chi} - E_R, ~-\vec{p}_R).
    \end{aligned}
\end{equation}

The final transformation to the lab frame is
\begin{equation}
    [p_2, p_3] = \text{T}1 \cdot R_{y}^{\pi - \theta} \cdot \text{T}2^{\dagger} \cdot \text{R}2^{\dagger}[p''^c_2, p''^c_3].
\end{equation}

\textit{Appendix C: EPA approximated cross sections} - To understand the dependency of the cross section on DM energy and lepton/mediator masses, we present the EPA approximations~\cite{Altmannshofer:2014pba, Belusevic:1987cw, Magill:2016hgc}, which have been used for neutrino tridents. As shown below, the $log$ dependence of lepton mass implies that the results can be extended to other $U(1)$ model scenarios.  
For $m_{Z'} \gg \sqrt{s}$,
\begin{equation}
    \sigma_{\chi,\gamma} = \frac{Z^2 \alpha_{em}^2}{18\pi^3} \bigg( \frac{g_{Z'}^2g_{D}^2}{m_{Z'}^4}\bigg) \bigg[ \log{\frac{\hat{s}}{m_l^2}} - \frac{19}{6} \bigg]
    \label{eq:EPAheavymz}
\end{equation}

For the regime where $m_l \ll m_{Z'} \ll \sqrt{\hat{s}}$
\begin{equation}
    \sigma_{\chi,\gamma} = \frac{1}{m_{Z'}^2}\frac{g_{Z'}^2g_{D}^2 \alpha_{em}}{6\pi^2}\log{\frac{m_{Z'}^2}{m_l^2}}
    \label{eq:EPAmediummz}
\end{equation}

For the regime where $m_{Z'} \ll m_{l} \ll \sqrt{\hat{s}}$
\begin{equation}
    \sigma_{\chi,\gamma} = \frac{1}{m_{l}^2}\frac{7 g_{Z'}^2g_{D}^2 \alpha_{em}}{72 \pi^2}\log{\frac{m_{l}^2}{m_{Z'}^2}}
    \label{eq:EPAsmallmz}
\end{equation}

One can obtain the total cross section by integrating the above between the limits as in Eq.~\eqref{eq:sq2limits}. 

The EPA approximated cross section approaches the analytical one as the virtuality of the photon decreases~\cite{Ballett:2018uuc} ($q^2\approx 0$). This occurs when (1) the incoming fermion is ultrarelativistic (which is always true for neutrino tridents) and (2) the peak momentum transferred is less than the mass of the final state leptons. For this reason, the above expressions have been obtained by setting $m_{\chi} = m_l = q^2 = 0$ in the matrix element. Since massive DM is not always ultrarelativistic, the EPA approximated cross section differs from the real one by at most one order of magnitude. 

\begin{figure}[h]
    \centering
    \includegraphics[width=0.46\textwidth]{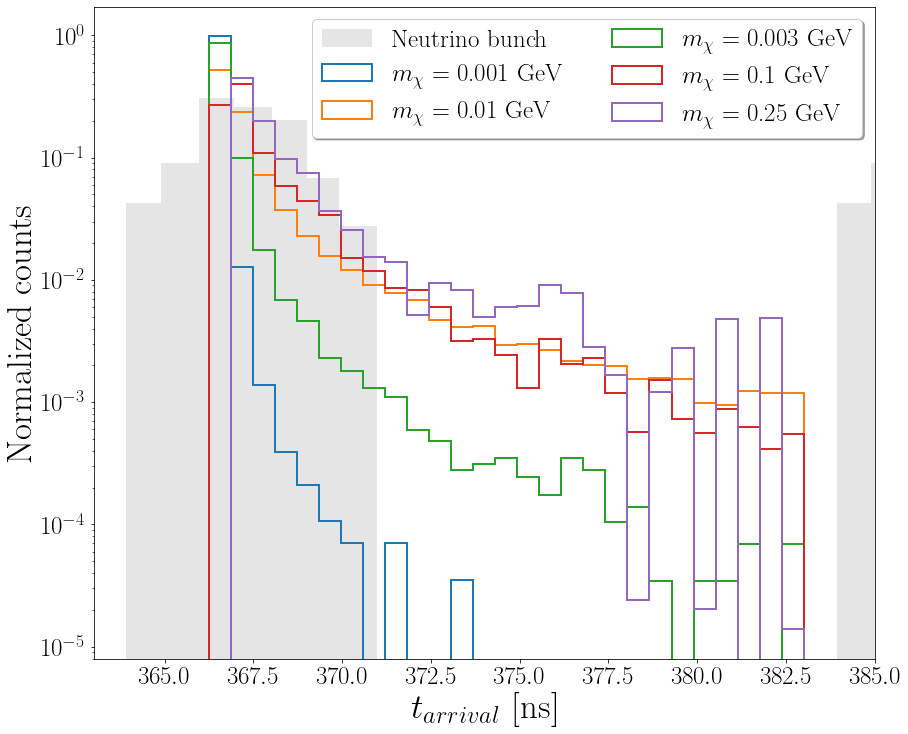}
    \captionsetup{justification=Justified, singlelinecheck=false}
    \caption{Spectra depicting arrival time of DM in comparison with that of neutrinos. }
    \label{fig:timingsbnd}
\end{figure}

\textit{Appendix D: Timing at SBND} -
Figure.~\ref{fig:timingsbnd} shows the time of arrival of the DM candidate at SBND in comparison with the true timing distribution (after accounting for the photon's time of flight to the PMTs) of SM neutrinos~\cite{SBND:2024vgn, bsmsbnd, bsmsbnd2} in one proton-beam spill. This study does not include a detailed simulation of the true timing distribution of the DM after accounting for the photon's time of flight, but we expect $\sim 2$~ns more delay in the DM timing spectra. We anticipate that even with $\sim2$~ns Gaussian smearing, we can achieve a background-free timing window between 371 and 380~ns. We reserve a detailed timing analysis for other DM models that can be probed at SBND.

\begin{figure}[h]
    \centering
    \includegraphics[width=0.46\textwidth]{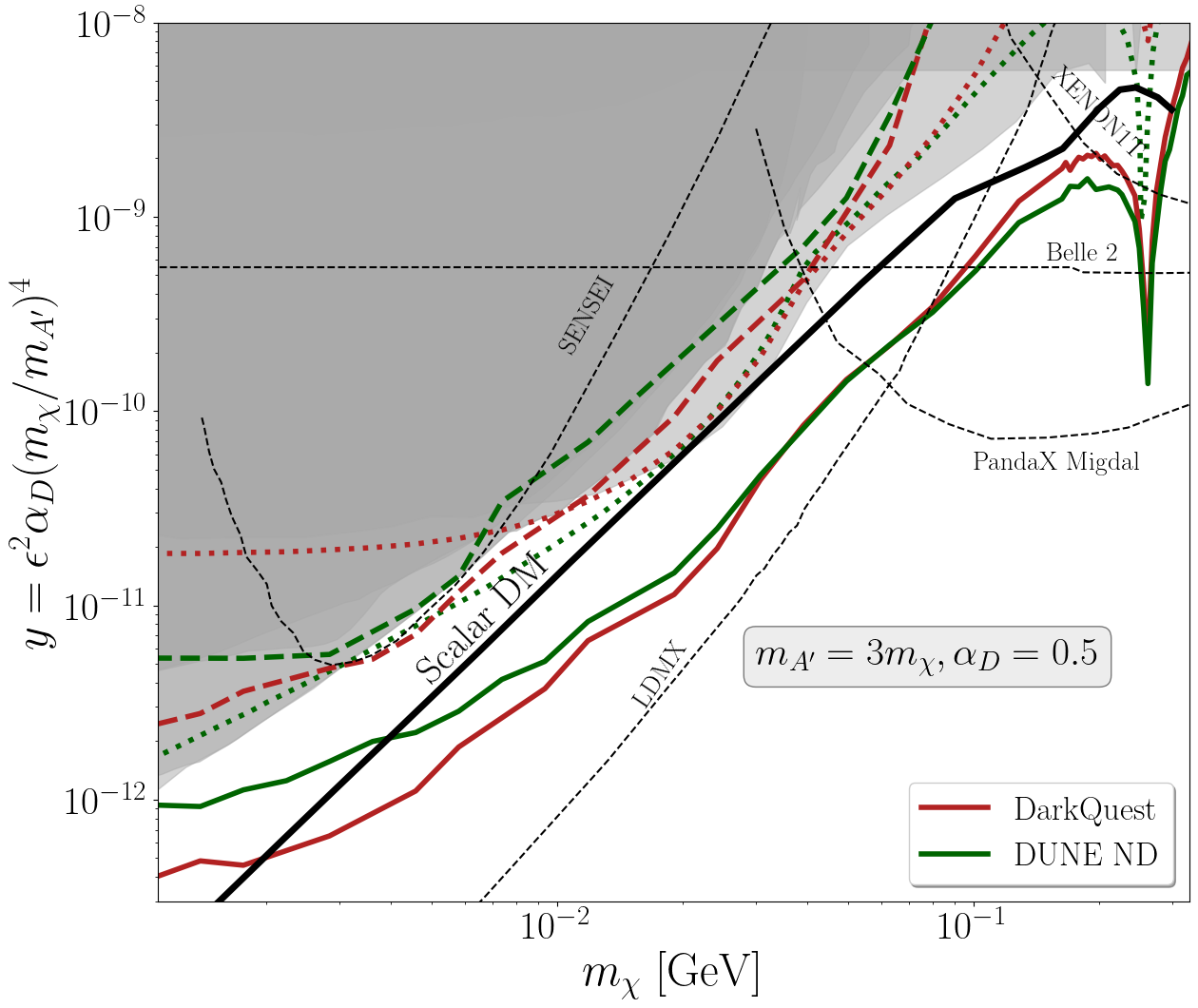}
    \captionsetup{justification=Justified, singlelinecheck=false}
    \caption{DIPP sensitivities (solid) in comparison with dark trident (dashed) and DM recoil (dotted) at DUNE ND (red) and DarkQuest Phase 2 (green)}
    \label{fig:DIPP_DT_DE}
\end{figure}

\textit{Appendix E: DIPP with versus dark trident and dark matter electron recoils} - Figure~\ref{fig:DIPP_DT_DE} shows the sensitivity of DM coupled to dark photons via DIPP, compared to dark tridents and DM-electron recoils. Although dark tridents share similar topologies ($l^+ l^-$), their sensitivity is suppressed by an extra $Z'$ mediator, introducing an additional $\epsilon^2/m_{A'}^2$ factor. Electron recoil sensitivities at DUNE ND are taken from Ref.\cite{Breitbach:2021gvv}. Since DarkQuest has not studied single-electron events, we estimate them using photon backgrounds from Ref.\cite{Blinov:2024gcw}, finding suppression due to large single-electron backgrounds. The figure also shows direct detection limits from SENSEI~\cite{SENSEI:2024yyt}, XENON1T~\cite{XENON:2019zpr}, and PandaX~\cite{PandaX:2023xgl}; the former two are excluded, while PandaX probes $m_\chi > 30~\text{MeV}$. Belle II (monophoton) can reach $y > 5 \times 10^{-9}$\cite{Inguglia:2019gub}, and LDMX phase I can access $y \gtrsim 10^{-12}$ for $m_{A'} \gtrsim 10$ MeV\cite{LDMX:2018cma}. These projections rely on DM–electron couplings, while PandaX probes quarks. By accessing both quark and lepton couplings, DIPP complements these searches and helps map the full parameter space.

\bibliography{references}

\end{document}